\documentclass[a4paper,11pt]{article}
\pdfoutput=1

\usepackage{jinstpub} 

\title{\Large{Evaluation of a method for time-of-flight, wavelength and distance calibration for neutron scattering instruments by means of a mini-chopper and standard neutron monitors}}

\author[a]{L. Vergara,}
\author[a]{M. Arai,}
\author[a]{R. Woracek,}
\author[a]{M. Olsson,}
\author[a]{A. Quintanilla,}
\author[a]{S. Alcock,}
\author[a]{J. Nilsson,}
\author[a]{K. Kanaki,}
\author[a]{P.M. Kadletz,}
\author[a,b]{O. Kirstein,}
\author[a,c,d]{R. Hall-Wilton}
\author[a,1]{and N. Tsapatsaris\note{Corresponding author.}}

\affiliation[a]{European Spallation Source ERIC (ESS) - P.O. Box 176, SE-22100 Lund, Sweden.}
\affiliation[b]{University of Newcastle, School of Mechanical Engineering,  NSW, Australia. }
\affiliation[c]{Università degli Studi di Milano-Bicocca, Piazza della Scienza 3, 20126 Milano, Italy.}
\affiliation[d]{Sensors and Devices Centre, Fondazione Bruno Kessler, Via Sommarive 18, 38123 Trento, Italy}

\emailAdd{nikolaos.tsapatsaris@ess.eu}

\abstract{Accurate conversion of neutron time-of-flight (TOF) to wavelength is of fundamental importance to neutron scattering measurements in order to ensure the accuracy of the instruments and the experimental results. Equally important in these measurements is the determination of uncertainties, and with the appropriate precision. Especially in cases where instruments are highly configurable, the determination of the absolute wavelength after any change must always be performed (e.g. change of detector position). Inspired by the manner with which neutron spectrometers determine the absolute wavelength, we evaluate for the first time, in the author's knowledge, a commonly used method for converting TOF to neutron wavelength, the distance of a monitor from the source of neutrons  and we analytically calculate the uncertainty contributions that limit the precision of the conversion. The method was evaluated at the V20 test beamline at the Helmholtz Zentrum Berlin (HZB), emulating the ESS source with a long pulse of 2.86~ms length and 14~Hz repetition rate, by using a mini-chopper operated at 140~Hz and two portable beam monitors (BMs), as well as accompanied data acquisition infrastructure. The mini-chopper created well-defined neutron pulses and the BM was placed at two positions, enabling the average wavelength of each of the pulses created to be determined. The used experimental setup resulted in absolute wavelength determination at the monitor positions with a $\delta \lambda_{mean} / \lambda_{mean} $ of $\sim$1.8\% for $\lambda >4$~Å. With the use of a thinner monitor, a $\delta \lambda_{mean} / \lambda_{mean} $ of $\sim$1\% can be reached and with a modest increase of the distance between the reference monitor positions a $\delta \lambda_{mean} / \lambda_{mean} $ of below 0.5\% can be achieved. Further improvements are possible by using smaller chopper disc openings and a higher rotational speed chopper. The method requires only two neutron measurements and doesn't necessitate the use of crystals or complex fitting with sigmoid functions and multiple free variables, and could constitute a suitable addition to imaging, diffraction, reflectometers and small angle neutron scattering instruments, at spallation sources, that do not normally utilise fast choppers.}


\begin{document}
\maketitle
\flushbottom

\section{Introduction}

The European Spallation Source (ESS), a European Research Infrastructure (ERIC), is a research facility under construction in Lund, Sweden, which will provide the world's most powerful neutron source yet \cite{Garoby_2018}. The facility is meant to enable scientific breakthroughs in several research fields, by reaching a higher flux and a longer neutron pulse length than the neutron sources of today. In order to serve the various scientific areas, 15 versatile neutron scattering instruments are planned for ESS  \cite{ANDERSEN2020instrumentSuite}. 

In order to fully utilise the long neutron pulse, all ESS instruments will be equipped with complex chopper systems and beam monitors (BMs), amongst other components \cite{TDRbook}. Neutron choppers are used to periodically interrupt the neutron beam for a well-defined duration, in order to structure and shape the pulse. They can therefore be used to produce the needed distribution of neutrons as a function of time \cite{First_choppers}. BMs are low-efficiency neutron detectors, which absorb a small percentage of the incoming beam, while simultaneously leaving the remaining beam undisturbed \cite{Vmon}. These monitors are used to measure the performance of the instruments, by recording the intensity of the beam as a function of time \cite{CharacterizationBeamMonitors}. Furthermore, in time-of-flight (TOF) experiments, BMs can be used to obtain the wavelength of pulses of neutrons; by placing two monitors at known positions in the beamline and recording the arrival time of the neutrons in both the monitors, the average wavelength $\lambda_{mean}$ of the neutrons can be found by:  

\begin{equation}
\lambda_{mean}=\frac{h(\text{tof}_{2,mean}-\text{tof}_{1,mean})}{mL},
\label{eq:tof}
\end{equation}
where $h$ is Planck's constant and $m$ is the mass of the neutron. $L$ is the distance between the two BMs, and tof is the time it takes for the neutrons to travel from the source to a monitor, i.e. the TOF of the neutrons. 
 
The conversion of time-of-flight to neutron wavelength is central to neutron scattering measurements, as accurate conversion of TOF to neutron wavelength ensures the accuracy of the cascade of calculations that lead for example in the determination of d-spacing in (poly)crystalline samples. Especially in cases where instruments are highly configurable, as will be the case at ESS, a re-calibration after any positional change of a detector must always be performed. One standard calibration method \cite{SantistebanBraggEdge} used in TOF neutron imaging involves the use of a standard polycrystalline Fe sample, where the most intense Bragg edges in the transmitted pattern are fitted individually in a least-squares refinement procedure. Other methods used at TOF small angle neutron scattering instruments at reactor sources utilise a fast scanning slit at the location of the sample and two different detector positions to determine the neutron wavelength and the d spacing of the diffracting planes of the reference samples \cite{SilverBehenate}. Although using Bragg edge scanning can be a very accurate method for converting TOF to neutron wavelength it necessitates the use of a powerful enough beam and short source pulses. It involves complex fitting of line shapes and the sigmoid function fitting must be done very carefully and in multiple stages (extremities and the edge separately) to avoid local fitting minima and often require collection of enough data to ensure fitting quality. In this work we evaluate a simple, reproducible calibration method for TOF instruments that allows the determination of the neutron wavelength and positions of monitors and detectors with respect to the neutron source without complex fitting, and introducing less and different systematic uncertainties than common methods. Furthermore, the method provides more data points in comparison to Bragg edges and there is no mosaic spread of crystals/powders inducing edge smearing and shifting, since no samples are utilized. The method was evaluated at the V20 beamline \cite{WORACEK,STROBL201374}, a dedicated ESS test beamline at Helmholtz Zentrum Berlin (HZB), and utilized a high-frequency mini-chopper together with a BM, which was placed at two different positions. Between 2015 and 2019, the European Spallation Source operated the V20 test beamline, at the 10~MW research reactor of HZB, which was utilized for the testing of choppers \cite{FORSTER2018298}, detectors \cite{Backis_2020} and other diverse topics \cite{Vmon, V20WFM_Lohmann}.

\section{Experimental setup} 

\begin{figure*}[h]
\begin{center}
\includegraphics[width=\textwidth]{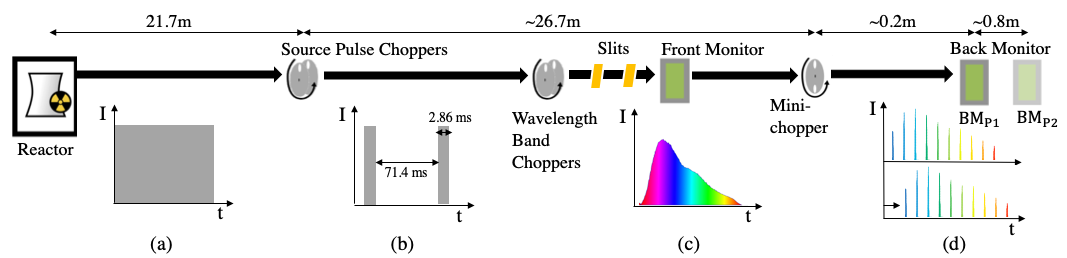}
\caption{V20 ESS test beamline instrument configuration at Helmholtz Zentrum Berlin and the development of the neutron stream from (a) the moderator, (b) directly after the Source Pulse Choppers, (c) before entering the mini-chopper (only one of the pulses from (b) is shown), and (d) after the mini-chopper, measured by a BM at two different positions, BM$_{\text{P1}}$ and BM$_{\text{P2}}$ (upper and lower series in the graph). The y-axes in the graphs represent intensity and the x-axes represent time.  The figure is not to scale.} 
\label{fig:cartoon}
\end{center}
\end{figure*} 

\begin{figure}[h]
\centering
\includegraphics[width=7cm]{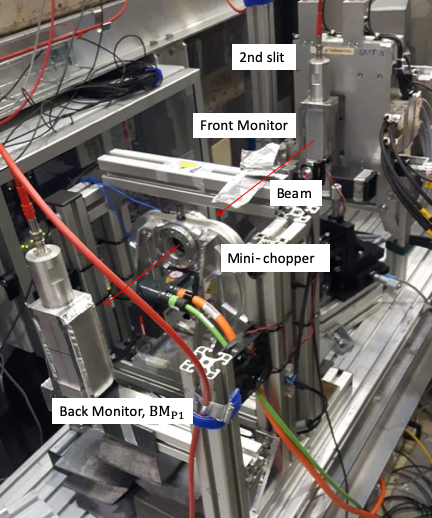}
\caption{Photo of the experimental setup at V20, where the mini-chopper can be seen approximately at the centre. }
\label{fig:setupPhoto}
\end{figure}

In figure~\ref{fig:cartoon} the configuration of the V20 test beamline at HZB, used for the evaluation of this method, can be seen as well as the modification of the neutron beam by the different choppers. A continuous stream of neutrons (see figure~\ref{fig:cartoon}a), originating at the reactor, are moderated and transported by neutron guides to the Source Pulse Choppers, a counter-rotating double chopper system. The Source Pulse Choppers are placed 21.75~m from the cold source and their objective is to shape the neutron beam to mimic the ESS neutron source, by creating pulses of approximately 2.86~ms of length at a repetition rate of 14~Hz (frame time 71.4~ms), as shown in figure~\ref{fig:cartoon}b. The neutrons contained within these pulses have different velocities and the Wavelength Band Choppers, which are placed after the Source Pulse Choppers, prevent frame overlap between the slowest neutrons of one pulse and the fastest neutrons of the succeeding one. Due to these different neutron velocities, a spread of the pulse (as in figure~\ref{fig:cartoon}c) is observed at the Front Monitor. The Front Monitor is a Multi Wire Proportional Chamber (MWPC) beam monitor, filled with $^3$He gas. It has an efficiency of ca. $10^{-5}$ at 1.8~Å and is produced by Canberra (former Dextray) of Mirion Technologies \cite{MIRION}. The neutron pulses then reach the mini-chopper (shown in figure~\ref{fig:setupPhoto}) operating at 140 Hz, which transmits a neutron velocity distribution defined by the openings of the mini-chopper  disc and its time offset in relation to the disc opening of the Pulse Source Choppers. This chopper consists of a 0.175 m diameter-disc with two equal openings \cite{ESSchopperPrototypeTestV20} of 6.4~mm at half the height of a 22~mm slit, giving a nominal angular opening of 4.8$^\circ$. After the mini-chopper, the neutron pulses ultimately reach the Back Monitor, which is the same type as the Front Monitor. It has, however, a higher efficiency of $3.9 \cdot 10^{-3}$ at 1.8~Å. The active thickness of the monitor is 40~mm. Although 
the efficiency of 
the BMs varies with wavelength, the band of wavelengths in this work is narrow, and the efficiency variation 
is therefore not relevant for the method described in this paper. 
Measurements were taken with the Back Monitor placed at two different positions (see figure~\ref{fig:cartoon}d), which were 0.7525~m apart, and took approximately 15~minutes per scan. The monitor positions downstream from the mini-chopper will, from hereon, be denoted by BM$_{\text{P1}}$ (closest to mini-chopper) and BM$_{\text{P2}}$. The Front Monitor will be denoted BM$_{\text{F}}$.

\subsection{Concept of operation}

\begin{figure}[h]
\centering
\includegraphics[width=8.5cm]{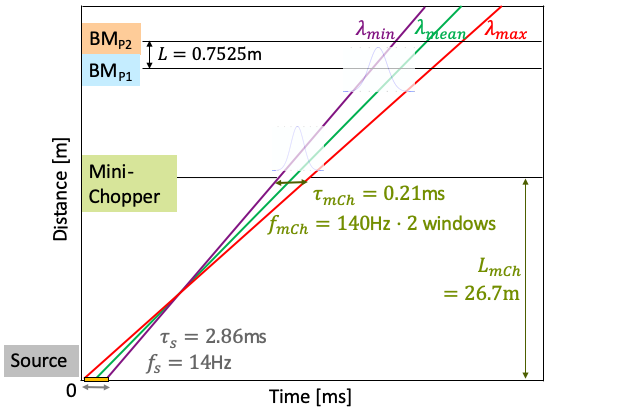}
\caption{TOF-diagram describing the trajectory of neutrons of lowest, mean, and highest wavelengths. }
\label{fig:TOFdiagram}
\end{figure}

In neutron scattering, the distance traveled by a neutron is usually shown as a function of time in time-of-flight diagrams, where the trajectory and velocity  of a neutron can be described by a straight line. The slope of the line corresponds to the velocity  of the neutron, where the steeper the slope, the faster the neutron. In the TOF-diagram shown in figure~\ref{fig:TOFdiagram} the trajectory of neutrons from the source to the back monitors' two positions can be seen. The neutron trajectories denoted $\lambda_{min}$, and $\lambda_{max}$, represent the fastest, and slowest, neutrons transmitted by the mini-chopper. The neutron trajectory denoted $\lambda_{mean}$ represents a neutron stemming from the nominal middle of the ESS source pulse, passing through the centre of the mini-chopper opening, and thus appearing at the centre of the pulses created by the mini-chopper. 

The wavelength range transmitted by the mini-chopper $\Delta \lambda_{mCh}$ is given by eq.~(\ref{eq:chopperLambdas}), where $\tau_s$ and $\tau_{mCh}$ are the opening times of the Source Pulse Choppers and of the mini-chopper, respectively, and $L_{mCh}$ is the distance between them. However, this definition of $\Delta \lambda$ is not used in this work, as the method relies only on measuring the average wavelength $\lambda_{mean}$ transmitted through the mini-chopper. This average wavelength is calculated by eq.~(\ref{eq:tof}), by only using the TOF at the centre of the pulses created by the mini-chopper. We assume, however, that the neutron transmission temporal profile is normally distributed, symmetric and that it can be defined by two limits ($\lambda_{min}$, $\lambda_{max}$). In the section \textit{Results and Discussion} it is discussed how precise the determination of the average wavelengths is. 

\begin{equation}
    \Delta \lambda_{mCh} = \lambda_{max} - \lambda_{min} = \frac{h(\tau_s + \tau_{mCh})}{mL_{mCh}}
    \label{eq:chopperLambdas}
\end{equation}

\subsection{Beam configuration} 

In order to avoid an artificial intensity variation due to the wavelength dependent angular divergence from the guide, the beam incident on the mini-chopper was collimated by two slits, located at a distance of 0.771~m from each other. The first slit was 14~mm~$\times$~10~mm of height and width, respectively, while the second was 11~mm~$\times$~10~mm. This configuration of the slits resulted in a theoretical FWHM beam width $BW_{theo}$ of
18.8~mm 
at the position of the mini-chopper disc, calculated by eq.~(\ref{eq:effBW})
\begin{equation}
    BW_{theo} = \sqrt{\left(\frac{d_{S2,CD}}{d_{S1,S2}}W_{S1}\right)^2 + \left(W_{S2} + \frac{d_{S2,CD}}{d_{S1,S2}}W_{S2}\right)^2},
     \label{eq:effBW}       
\end{equation}
where $W_{S2}$ is the width of the second slit, $W_{S1}$ the width of the first slit, $d_{S1,S2}$=0.771~m is the distance between the first and second slit, and $d_{S2,CD}$=0.563~m is the distance between the second slit and the chopper disc. 

For a single disk, the burst time of an emitted neutron pulse is defined as the total time the chopper disc opening overlaps with the beam width, i.e. the time of any neutron transmission. Given that the angle of the disc opening is smaller than the angle of the equivalent beam width, the burst time (eq.~(\ref{eq:burstTsum})) is the sum of (i) the time that the chopper disc opening starts to  overlap with the beam (only partial overlap) $t_{start}$, (ii) the time the opening and the beam fully overlap $t_{plateau}$, and (iii) the time it takes the opening to end its overlap with the beam (only partial overlap again) $t_{end}$. 
\begin{equation}
    t_{burst} = t_{start} + t_{plateau} + t_{end}
    \label{eq:burstTsum}
\end{equation}
\begin{equation}
    t_{burst} = (\alpha_{CO} + (\alpha_{BW}-2\alpha_{CO}) + \alpha_{CO})\frac{T}{360}
    \label{eq:burstT}
\end{equation}
The burst time was calculated by eq.~(\ref{eq:burstT}), where $T$($=1/f$) is the period of the chopper, $\alpha_{BW}$ is the angle corresponding to the effective beam width and $\alpha_{CO}$ is the disc chopper opening angle.

The angle $\alpha_{BW}$ can be calculated by eq.~(\ref{eq:BWangle}), where $BW$ 
is the
beam width, $r_{CD}$ is the radius of the chopper disc and $h_{CO}$ is the height of the chopper disc opening.
\begin{equation}
    \alpha_{BW} =  \frac{BW_{theo}}{2\pi (r_{CD}-h_{CO}/2)} \cdot 360^\circ
    \label{eq:BWangle}
\end{equation}
The theoretical FWHM beam width BW$_{theo}$, calculated with eq.~(\ref{eq:effBW}), was used to calculate $\alpha_{BW}$, yielding a result of 14.08$^\circ$, which in turn yielded a burst time of 279.37~$\mu$s. This burst time was then used to calculate a theoretical FWHM, as in eq.~(\ref{eq:theoFWHM}), where $t_{start}$ and $t_{end}$ were also calculated by eq.~(\ref{eq:burstT}). 
\begin{equation}
    \text{FWHM}_{theo} = t_{burst} - \frac{t_{start}}{2} - \frac{t_{end}}{2}
    \label{eq:theoFWHM}
\end{equation}
The theoretical FWHM of each emitted pulse was found to be 184.12~$\mu$s. Although the beam's horizontal dimensions were not ideal for creating a short burst time, the low neutron flux didn't allow for further collimation of the beam, as it would have resulted in a significant loss of neutrons. 

\subsection{Chopper control system and data acquisition} 

The mini-chopper controller is based on a Beckhoff PLC (Programmable Logic Controller \cite{PLC}). The controller is divided into two cpu cores, which control (i) the phase and the speed of the chopper and (ii) the communications with EPICS (Experimental Physics and Industrial Control System \cite{EPICS}), through ModbusTCP/IP \cite{TCPIP}, by translating the data from the chopper to a standardized language for EPICS.  The mini-chopper drive terminal is an EL7211 and the motor  is a Beckhoff AM8121. Additionally, the chopper controller verifies the correct chopper function through condition monitoring.  
 
All chopper drives at the V20 instrument are controlled through a common timing system reference signal. Furthermore, each chopper disc revolution is time stamped using a rotation sensor \cite{ESSVerticalIntegration}. The timing system gathers information from EPICS and an EPICS forwarder \cite{Mukai_2018} delivers information to Apache Kafka \cite{kafka}
. Apache Kafka also receives information from the BMs; the amplified signal of the BMs is connected to the Data Acquisition System (DAQ), a four channel system that digitizes the input. The data is then forwarded to the Event Formation Unit (EFU), before being delivered to Apache Kafka, which then directly provides information to the NeXus file writer, that saves the data to disc. A detailed description of EPICS, the DAQ and timing systems, as well as the mini-chopper control system and integration at V20 can be found in \cite{ESSchopperPrototypeTestV20}.

\section{Results and discussion}

\begin{figure}[h]
\centering
\includegraphics[width=8.5cm]{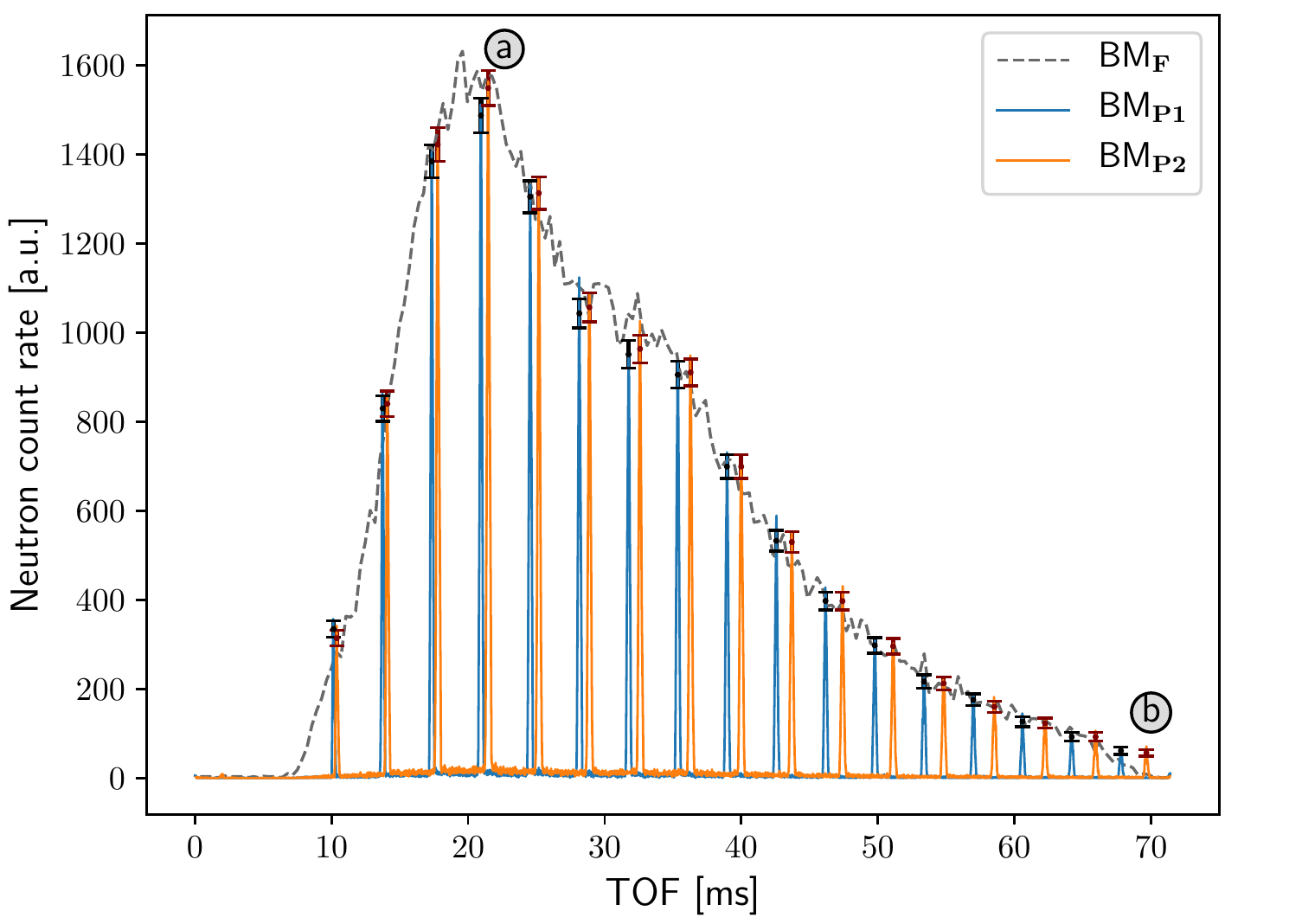}
\caption{The neutron count rate as a function of TOF, measured by the Front Monitor BM$_{\text{F}}$ and by the Back Monitor, at two different positions: BM$_{\text{P1}}$ and BM$_{\text{P2}}$. Statistical uncertainties are shown, for clarity, only at the peak of each pulse. The pulses denoted a and b are shown again in figure~\ref{fig:fittedPeaks}.} 
\label{fig:140HzOverlay}
\end{figure}

Each pulse, the Source Pulse Choppers transmit, spreads-out due to the different neutron velocities contained within and finally arrives at the front monitor BM$_{\text{F}}$ as shown in figure~\ref{fig:140HzOverlay} (compare with figure~\ref{fig:cartoon}c).  

The noise level of the data evident in the Front Monitor measurement BM$_{\text{F}}$ is significantly higher compared to the Back Monitor (BM$_{\text{P1,2}}$) due to the lower efficiency of the available diagnostic Front Monitor and therefore lower count rate. The blue and orange traces in figure~\ref{fig:140HzOverlay} represent the neutron pulses created 
by the mini-chopper, which produced two well-defined pulses per revolution (compare with figure~\ref{fig:cartoon}d). Due to the curved guide of the V20 instrument no neutrons with wavelength below 1.1~Å are transmitted. The neutron count rate was measured by the Back Monitor at two different positions in order to determine the average flight time of the neutrons between these positions. The $1/\lambda$--efficiency of the monitors is not taken into account, as the absolute intensity does not change significantly for pulses of similar wavelength and is not utilised in this analysis. 
As can be seen in figure~\ref{fig:140HzOverlay}, the incident spectrum coincides well with the spectra measured by the Back Monitor.

\begin{figure}[h]
\centering
\includegraphics[width=10cm]{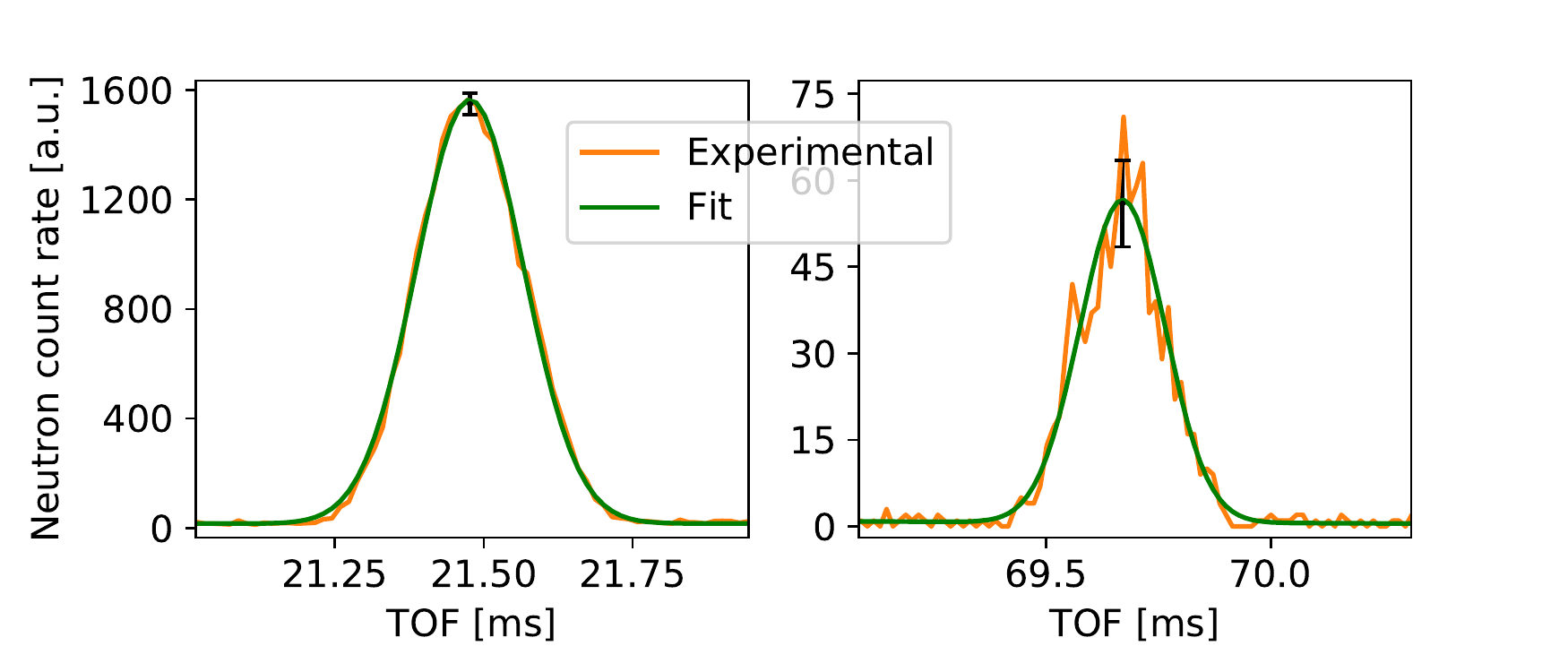}
\caption{Individually fitted pulses created by the mini-chopper fitted using a 
Gaussian curve and a linear background for the TOF determination. The pulses in the figure are from the second monitor position, at \textit{a} and \textit{b} in figure~\ref{fig:140HzOverlay}.}
\label{fig:fittedPeaks}
\end{figure}

In order to calculate the average wavelength  of each of the pulses created, the TOF at the centre of the pulses ($\text{tof}_{1,mean}$ and $\text{tof}_{2,mean}$ in eq.~(\ref{eq:tof})) had to be determined. For this TOF-determination a 
Gaussian curve and a linear background were fitted to each pulse individually. Two fitted pulses from the second monitor position (pulses a and b from  figure~\ref{fig:140HzOverlay}) can be seen in figure~\ref{fig:fittedPeaks}. The fitting quality is very high for the lower-TOF pulses (left), for both monitor positions, but declined at higher TOF (right), as the neutron count-rate decreased due to a lower flux at longer wavelengths. 

\begin{figure}[h]
\centering
\includegraphics[width=8.5cm]{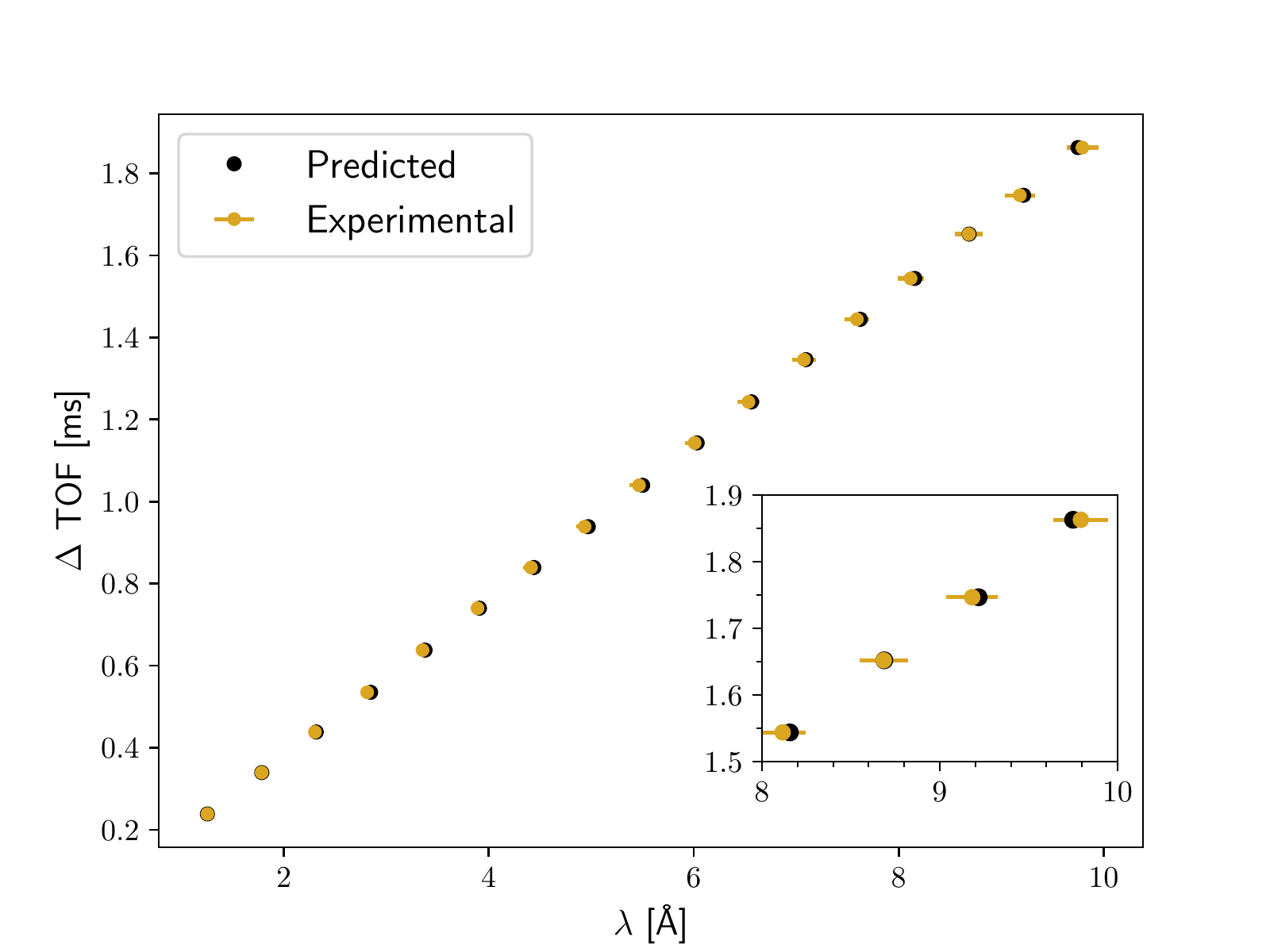}
\caption{The experimentally calculated and predicted wavelengths, vs. the TOF between the two monitor positions. }
\label{fig:LcalcANDpred}
\end{figure}

Once the TOF-values were extracted from the fit, eq.~(\ref{eq:tof}) was used to calculate the corresponding average wavelengths shown in the orange trace of figure~\ref{fig:LcalcANDpred}. In order to validate the calculations and that the mini-chopper was functioning correctly, the average wavelengths were also calculated by only taking into account the arrival of the first pulse and calculating the arrival of all subsequent pulses. For this, eq.~(\ref{eq:tof2Lpredicted}) was used:

\begin{equation}
    \lambda_{n,pred}= \frac{h(\text{tof}_{P2,0} + nt_{dP2}) - h(\text{tof}_{P1,0} + nt_{dP1} )}{mL}
    \label{eq:tof2Lpredicted}
\end{equation}
where tof$_{P1,0}$ and tof$_{P2,0}$ are the TOFs of the first pulse arriving at the first and second monitor positions, respectively, and $n$ is the number of the pulse, with $n=0$ for the first pulse. The variables $t_{dP1}=3.598$~ms and $t_{dP2}=3.699$~ms correspond to the time differences of the first and second pulse arriving at each monitor position. Figure~\ref{fig:LcalcANDpred} demonstrates a satisfactory agreement between the calculated average wavelengths extracted from the measured TOF-values, and predicted average wavelengths (from projected TOF-values), calculated by eq.~(\ref{eq:tof2Lpredicted}).  
The inset of figure~\ref{fig:LcalcANDpred} shows the last four points of the main figure, where some disagreement with theory can be seen due to the accumulation of uncertainties as the number $n$ of the pulses increases. 

\begin{figure}[h]
\centering
\includegraphics[width=8.5cm]{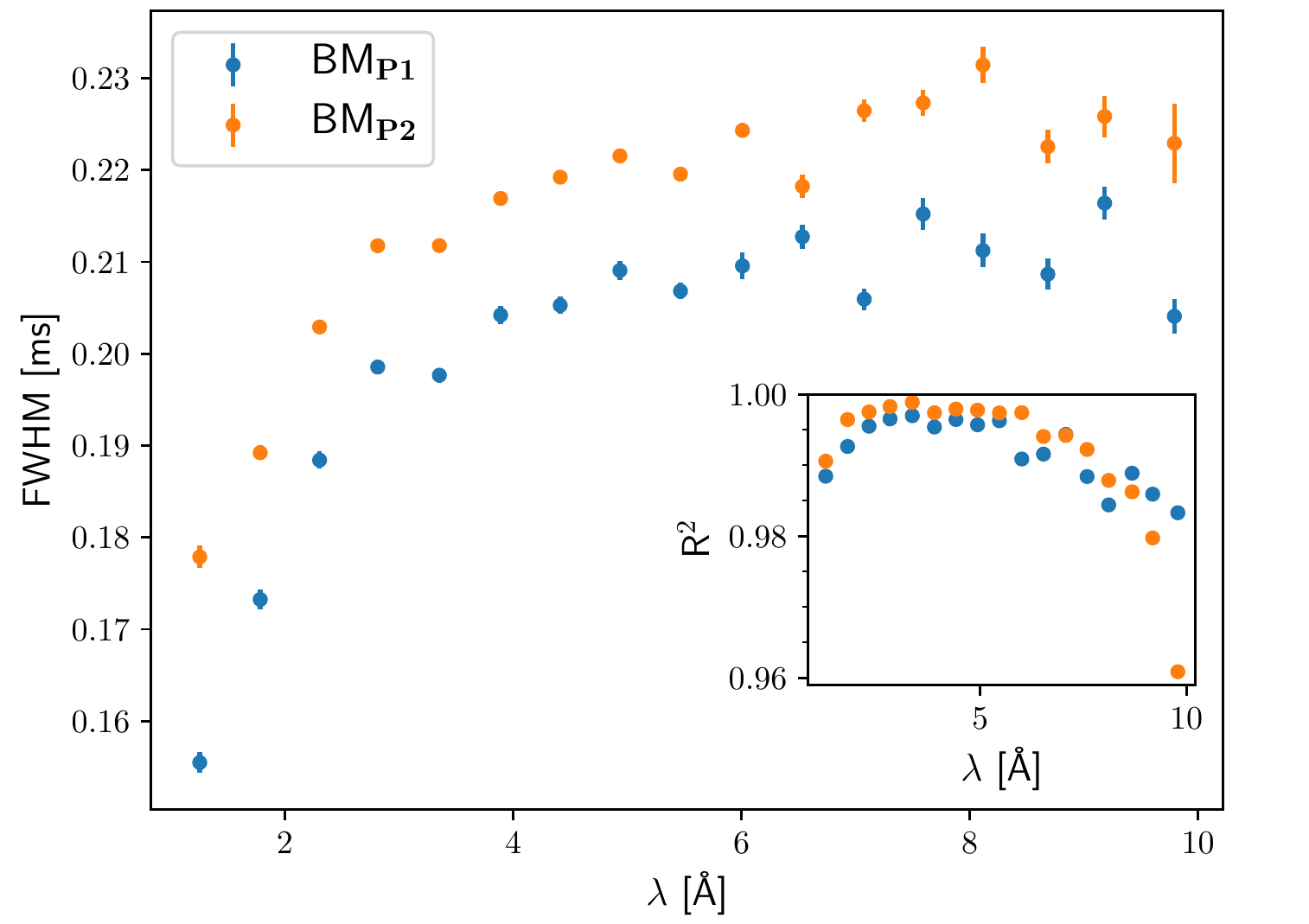}
\caption{FWHM values of each of the fitted pulses for BM$_{\text{P1}}$ and BM$_{\text{P2}}$ monitor positions. The inset shows the R$^2$ goodness of fit.}
\label{fig:FWHM}
\end{figure}

The FWHM of each pulse transmitted by the mini-chopper was obtained from the Gaussian fits, and is shown in figure~\ref{fig:FWHM} as a function of the experimental wavelengths. The lower-$\lambda$ pulses have a smaller width due to optical guide effects, which result in a lower divergence of the beam and hence, in narrower pulses. Furthermore, the FWHM is larger for all pulses of the second monitor position because of the spreading of the neutron pulses containing neutrons of different energies. Overall, the average measured FWHM of 205~$\mu$s at BM$_{\text{P1}}$  
is in acceptable agreement with the theoretically calculated FWHM of 184~$\mu$s. The disagreement with the experimentally measured FWHM is attributed to the limited transmitted neutron divergence from the neutron guide to the first slit. As a test of consistency, Eqs.~\ref{eq:burstT} and \ref{eq:BWangle} were combined, solved for $BW_{theo}$
 -- and using a burst time of 300.2~$\mu$s, obtained from a FWHM of 205~$\mu$s (from eq.~(\ref{eq:theoFWHM})) -- an experimentally derived FWHM beam width of 20.2~mm was obtained, which again compares well with the analytically derived 18.8~mm. The inset in figure~\ref{fig:FWHM} shows the R$^2$ goodness of fit measure for all the pulses, with $R^2$ above 0.99 up to a wavelength of 7.5~Å. 
 
\subsection{Determination of uncertainties}

In order to estimate the uncertainty in the determination of the average wavelengths, the general rule for uncertainties in a function of several independent variables, eq.~(\ref{eq:errorProp}), was used. The uncertainties that were propagated were categorized into TOF-- and distance.
\begin{equation}
    \delta f = \sqrt{\left(\frac{\partial f}{\partial x}\right)^2 (\delta x)^2 + \left(\frac{\partial f}{\partial y}\right)^2 (\delta y)^2 + ... }
    \label{eq:errorProp}
\end{equation}

The uncertainties that contributed to the determination of the TOF, and hence to the total TOF-uncertainty $\delta t_{tot}$, were the estimated uncertainties of the TOF-values (given by the fit) $\delta t_{1}$ and $\delta t_{2}$, as these variables are directly included in the calculation of the wavelength. The $\delta t$ differed for every tof-value, with a range between 0.2~$\mu$s and 0.9~$\mu$s. The mini-chopper has a phase uncertainty $\delta$phase of $\pm$0.15~$\mu$s, which also adds an uncertainty in the determination of the TOF-values. Moreover, the TOF-determination is also affected by the bin-size that is selected to display the collected data \cite{WhereToStickYourDataPoints}. The bin-size of our data points is 14~$\mu$s, and assuming a monotonically increasing/decreasing normal distribution of data the bin-size uncertainty $\delta$bin is one standard deviation or $\pm$4.76~$\mu$s. Although this uncertainty is one-sided, in this work we assume a symmetric uncertainty, due to its small contribution. 
The equation for calculating $\delta t_{tot}$ 
was derived, from eq.~(\ref{eq:errorProp}), to be eq.~(\ref{eq:totTimeErr}). 

One of the uncertainties that contribute to the total distance uncertainty $\delta d_{tot}$ stems from  
the measurement of the distance between the two monitor positions $\delta L$, which is estimated to be 0.5~mm, due to a cautious measurement by a calibrated ruler. The second uncertainty that contributes to $\delta d_{tot}$ 
is the gas-thickness \cite{Backis_2020} of the Back Monitor, $\delta$gas
. It is assumed in this test that the neutron detection is equally probable throughout the dilute gas; the gas has a thickness of 40~mm, which would give a maximum possible uncertainty of $\pm20$~mm. Converting this to a standard uncertainty by dividing it with $\sqrt{3}$, according to \cite{NISTrectangularDistr}, gives a $\delta$gas of 11.5~mm.
The total distance uncertainty is then calculated by eq.~(\ref{eq:totDistErr}) (derived by eq.~(\ref{eq:errorProp})) to be 11.51~mm. 

\begin{figure}[h]
\centering
\includegraphics[width=8.5cm]{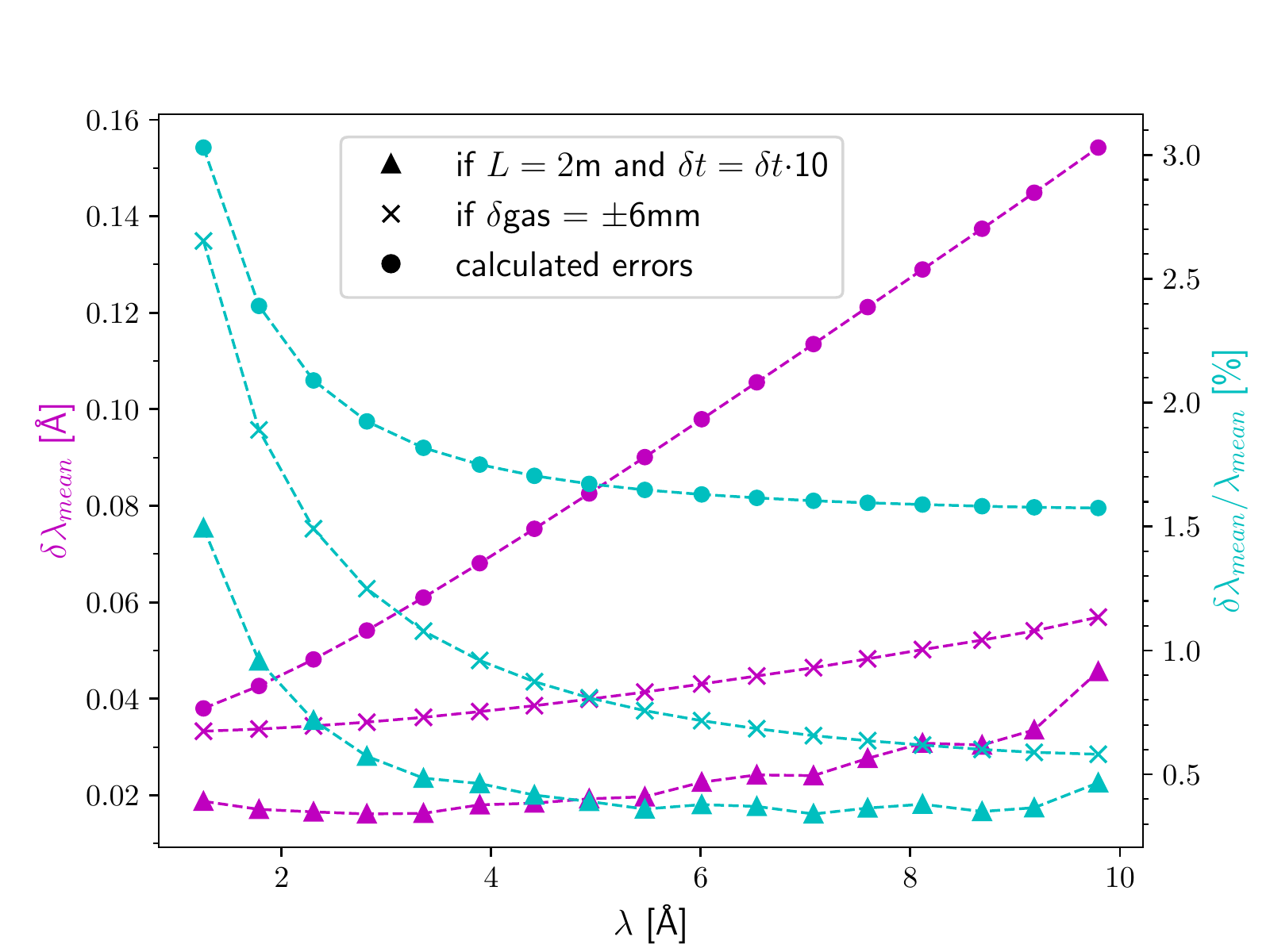}
\caption{The average wavelength uncertainty at the centre of the pulses (magenta) and average wavelength resolutions (turquoise) from actual experimental setup ($\bullet$) and hypothetical cases: ($\boldsymbol{\times}$) a thinner BM which provides a maximum $\delta$gas of 3.5~mm 
instead of 11.5~mm and ($\blacktriangle$) an increased distance between the two monitor positions together with a tenfold increase of the uncertainties of the TOFs.}
\label{fig:LambdaErr}
\end{figure}

Eq.~(\ref{eq:errorProp}) was finally used to derive eq.~(\ref{eq:lambdaErr}), the equation for the determination of the average wavelength uncertainties at the centre of the pulses $\delta \lambda_{mean}$, where $\Delta$tof$\, =\text{tof}_{2,mean}-\text{tof}_{1,mean}$. 
\begin{equation}
    \delta t_{tot} = \sqrt{(\delta t_2)^2 + (\delta t_1)^2 + 2(\delta \text{bin})^2 + 2(\delta \text{phase})^2}
    \label{eq:totTimeErr}
\end{equation}

\begin{equation}
    \delta d_{tot} = \sqrt{(\delta L)^2 + (\delta \text{gas})^2}
    \label{eq:totDistErr}
\end{equation}

\begin{equation}
    \delta \lambda_{mean} = \sqrt{\frac{h^2}{m^2L^2} (\delta t_{tot})^2 + \frac{h^2(\Delta\text{tof})^2}{m^2L^4}(\delta d_{tot})^2 }
    \label{eq:lambdaErr}
\end{equation}
In figure~\ref{fig:LambdaErr} the calculated ($\bullet$) average wavelength uncertainty $\delta \lambda_{mean}$ and the resolution of the average wavelength uncertainty 
$\delta \lambda_{mean} / \lambda_{mean}$ can be seen as a function of $\lambda$. In the same figure hypothetical uncertainties and resolutions are shown, which consider two cases: ($\boldsymbol{\times}$) a thinner BM with an active thickness of 12~mm instead of 40~mm, providing a maximum $\delta$gas of 3.5~mm 
instead of 11.5~mm
, and ($\blacktriangle$) a distance between the two monitor positions of 2~m (instead of 0.7525~m) and tenfold increase of the uncertainties of the TOFs caused by the worsening of the Gaussian fit quality due to the dispersion of neutron pulses by the time the neutron pulses arrive at the 2~m position.  The number of counts would also, unless a sufficiently large monitor is used, decrease by 40\%, due to the solid angle. Although the calculated $\delta \lambda_{mean} / \lambda_{mean}$ is a reasonable starting point for many instruments at ca.~
1.75\% for $\lambda>4$~Å, it can be significantly improved and reach approximately below 0.5\% by a more careful selection of the distance between the two positions of the Back Monitor.

\subsection{Conversion to distance}

\begin{figure}[h]
\centering
\includegraphics[width=8.5cm]{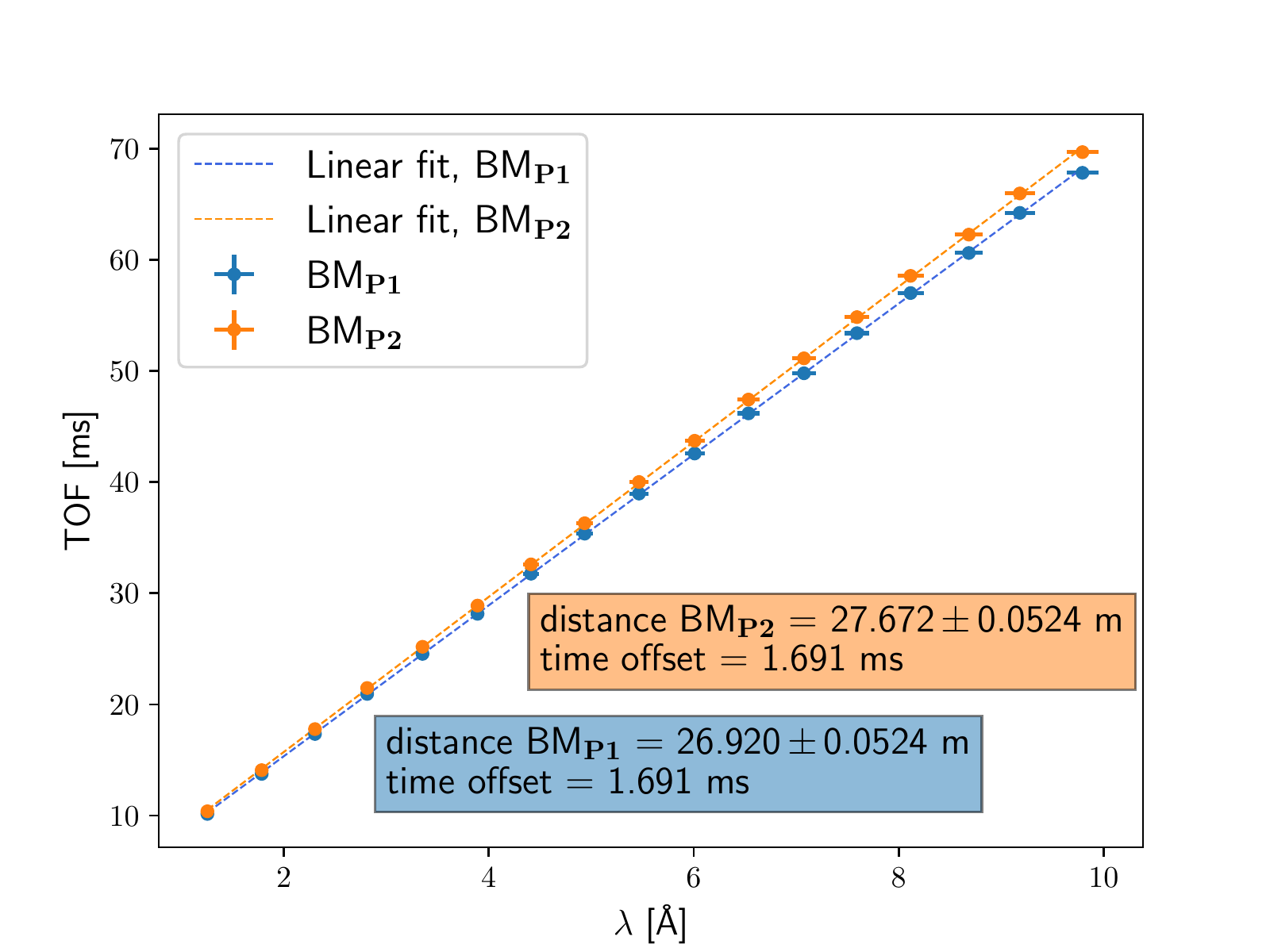}
\caption{The calculated wavelengths vs. the TOF from the neutron source to the two monitor positions. }
\label{fig:linearfitsVSlambda}
\end{figure}

Once the average wavelength of the pulses was determined, the two distances from the Back Monitor to the virtual neutron source, i.e. the Source Pulse Choppers, were extracted. The TOFs to the two monitor positions were plotted as a function of the calculated average wavelengths and a linear 
least-squares fit was made, as can be seen in figure~\ref{fig:linearfitsVSlambda}. The slope $a$ of the linear fits allows the calculation of the distance of the corresponding monitor from the source, as can be seen in Eqs.~\ref{eq:linear} and \ref{eq:distance}: 
\begin{equation}
    \text{tof} = a\lambda + t_{\text{offs}} \implies a = \frac{\Delta \text{tof}}{\Delta \lambda}
    \label{eq:linear}
\end{equation}
\begin{equation}
    L_S =\frac{h}{m}\frac{\text{tof}}{\lambda} = \frac{h}{m}a.
    \label{eq:distance}
\end{equation}

Eq.~(\ref{eq:linear}) is the linear equation, with t$_{\text{offs}}$ being the time offset, and the equation for the slope of a line. eq.~(\ref{eq:distance}) is derived from eq.~(\ref{eq:tof}), with $L_S$ being the distances from the back monitor to the neutron source. Time offset is the systematic delay caused by a fixed offset at the Source Pulse Chopper and in the acquisition electronics of the monitor. 
The distances to the monitor positions BM$_{\text{P1}}$ and BM$_{\text{P2}}$ were calculated to be 26.9197$\,\pm\,$0.0524~m and 27.6722$\,\pm\,$0.0524~m (linear fit uncertainty). This gives a difference of 0.7525~m between the BM$_{\text{P1}}$ and BM$_{\text{P2}}$ position, which confirms the original hand measurement. Deducing from the engineering design in \cite{WORACEK} BM$_{\text{P1}}\approx$ 26.983~m and BM$_{\text{P2}}\approx$ 27.736~m to the first and second monitor positions, respectively. Hence, the difference of the calculated distances versus the engineering design are 6.3~cm and 6.4~cm. The calculated time offsets t$_{\text{offs}}$ was  1.69~ms for both monitor positions, indicating that the neutron pulses originate from the same average position in time within the source pulse (green trace in figure~\ref{fig:TOFdiagram}).

\section{Conclusion and outlook}

In this work a method for (i) converting neutron TOF to wavelength and (ii) monitor distance to neutron source, using a 140~Hz neutron mini-chopper together with a beam monitor, was evaluated. The method is easy to reproduce and requires only two neutron scattering measurements. Moreover, the method does not involve crystals or complex fitting. The used experimental setup resulted in a $\delta \lambda_{mean}/ \lambda_{mean}$ of
1.75\% for $\lambda >4$~Å, a reasonable starting point for many instruments. Furthermore, we estimate that by increasing the distance between BMs a $\delta \lambda_{mean} / \lambda_{mean}$ of below 0.5\% can be reached. The resolution of the average wavelength can be improved further by using a thinner neutron monitor, smaller chopper disc openings and a higher speed chopper. The method will be further tested and compared to existing crystal-based methods at ESS and will complement the calibration arsenal of methods with different systematic uncertainties that ensures better calibration for neutron scattering instruments at spallation sources. With a $\delta \lambda_{mean} / \lambda_{mean}$ of below 0.5\% the method could, in certain cases, even be used as a replacement to crystal-based methods.

\end{document}